\documentclass[aps,prb,twocolumn,showpacs]{revtex4}
\usepackage{epsfig}
\usepackage{graphicx}

\begin{document}
\title{Superconductor-insulator transition driven by local dephasing}
\author{M.~Cuoco$^{(a,b)}$ and J.~Ranninger$^{(a)}$}
\affiliation{${(a)}$ Centre de Recherches sur les Tr\`es Basses
Temp\'eratures
 associ\'e \`a l'Universit\'e Joseph Fourier,
 C.N.R.S., BP 166, 38042 Grenoble-C\'edex 9, France}

\affiliation{${(b)}$I.N.F.M. Udr di Salerno, Laboratorio Regionale
SUPERMAT, Baronissi (Salerno), Italy \\
Dipartimento di Fisica ``E. R. Caianiello'', Universit\`a di
Salerno, I-84081 Baronissi (Salerno), Italy}

\begin{abstract}
We consider a system where localized bound electron pairs form an
array of "Andreev"-like scattering centers and are coupled to a
fermionic subsystem of uncorrelated electrons. By means of a
path-integral approach, which describes the bound electron pairs
within a coherent pseudospin representation, we derive and analyze
the effective action for the collective phase modes which arise
from the coupling between the two subsystems once the fermionic
degrees of freedom are integrated out. This effective action has
features of a quantum phase model in the presence of a Berry phase
term and exhibits a coupling to a field which describes at the
same time the fluctuations of density of the bound pairs and those
of the amplitude of the fermion pairs. Due to the competition
between the local and the hopping induced non-local phase dynamics
it is possible, by tuning the exchange coupling or the density of
the bound pairs, to trigger a transition from a phase ordered
superconducting to a phase disordered insulating state. We discuss
the different mechanisms which control this occurrence and the
eventual destruction of phase coherence both in the weak and
strong coupling limit.

\end{abstract}
 \maketitle


\section{Introduction}

The problem of interacting Cooper pairs and/or bosons, together
with the possibility of a quantum control of long range phase
coherence in low dimensional systems has received considerable
interest since the experiments\cite{Hav89} on homogeneous lead and
bismuth films, which exhibit a transition from a superconducting
to an insulating phase as a function of thickness.

It is known, since the work of Abrahams et al. \cite{Abr79}, that
no true metallic behavior can be expected to be observed for 2D
non-interacting electrons at T=0, because all the states will be
localized by an arbitrarily small amount of disorder. When one
includes Coulomb interaction, the situation is less clear, but the
common belief is that a metallic phase should still not appear at
T=0 in the presence of disorder - though no rigorous proof is
available\cite{Lee85}. Yet, in the presence of attractive
interactions, one expects a superconducting state both at T=0 and
finite temperature, even in presence of a finite amount of
disorder, due to its relative ineffectiveness for a transition of
Kosterlitz-Thouless type (the degree of relevance being given by
the Harris criterion\cite{Harris}).

The existence of a superconducting state at T=0 in 2D systems is
then considered to be directly linked to that of an insulating
state, with no intermediate metallic phase present. One hence
should  be able to observe a direct superconductor-insulator
transition (SIT) at zero temperature in 2D as a function of
disorder, interaction strength, magnetic field or any other
external parameter which can drive the system away from the
superconducting phase.

There are different theoretical scenarios which are commonly
discussed in connection with the salient features of such quantum
phase transitions: i) dissipative models, considering a network of
Josephson coupled superconducting grains shunted by
resistors\cite{Efe80,Ambe82,Lark83,Eck84,Cha87,Eme95}, where the
competition between the Josephson intergrain coupling and the
charging energy yields an increase of the phase fluctuations of
the superconducting order parameter and hence leads to a phase
disordered state. Dissipation thereby plays the role of
suppressing quantum phase fluctuations and thus competes with the
charging mechanism. ii) Bose-Hubbard models, where the
superconducting phase is due to the Bose-Einstein condensation of
charge-2e bosons and the insulating phase due to a proliferation
of vortices and localization of
pairs\cite{Fish89,Fish90,Kamp93,Sor92}.

While the SIT was long thought to be a paradigm of the above
theories, recent experiments have thrown some doubt about that.
They revealed, what seems to be a low-temperature metallic state
which is intertwined in such a SIT and thus requires a more
appropriate theoretical description. In magnetic field tuned
experiments on Mo-Ge samples\cite{Eph96}, in granular
superconductors (Ga films\cite{Jae86} and Pb films\cite{Ger90}),
and in Josephson junction arrays\cite{Zan92} a metallic phase has
been observed. Moreover, the recent experiments of Kapitulnik et
al.\cite{Kap99}, in which a metallic phase has been observed
sandwiched between the superconductor and the insulating phase,
suggest that perhaps two phase transitions accompany the loss of
phase coherence in 2D superconductors : (i) a superconductor to a
``Bose metal'' and (ii) a ``Bose metal'' to an insulator. A ``Bose
metal'' in this context, is thought of just a gapless
non-superfluid liquid with metallic like transport\cite{Phil03}.

These experimental results have led to reconsider the whole issue
of the SIT. One recent proposal to handle such a new viewpoint of
the SIT has been to reexamine the standard on-site charging model.
By including  nearest neighbor charging terms it was shown that
the resulting uniform Bose metal state lacks any trace of either
phase or charge order\cite{Das99}, due to a competition between
the order parameters which describe the onset of charge order and
that of phase coherence. A different point of view\cite{Dav99} is,
that in the quantum disordered regime a cancellation happens
between the exponentially long quasi-particle scattering time and
the exponentially small quasi-particle population, which
ultimately leads to a finite dc conductivity. Finally, intrinsic
as well as extrinsic sources of dissipation have been suggested as
potentially relevant for the occurrence of a non-superfluid
metallic phase in proximity to the superconducting
phase\cite{Kap99,Wag97,Rim97}. In this context, the dissipation is
a relevant perturbation\cite{Sac98}, which, depending on the
strength of the coupling to the dissipative source, can drive the
system from a {\it superconductor-to-insulator transition} to a
{\it superconductor-to-metal transition}\cite{Kap99,Kap01}.
Several other ideas supported a scenario where the system could
break into superconducting and insulating "islands" weakly linked
via percolating paths before going into a metallic phase,
dominated by vortex dissipation\cite{Kap01}. This "puddle"
scenario matches with the view to describe the 2D
superconductor-metal transition via superconducting islands
embedded in thin metal films\cite{Feig98}. Detailed analysis of
the puddle-like model considered strongly fluctuating
superconducting grains embedded in a metallic matrix\cite{Spiv01},
predicting a {\it metal-to-superconductor transition} with a
metallic phase just above the transition dominated by Andreev
reflections between the superconducting grains. Finally, recent
investigations directed  the attention to new phases with Bose
metallic features, where the dissipation is dynamically
self-generated, as in the quantum phase glass model, where
disorder in the distribution of the tunnelling amplitudes and
quantum fluctuations destroy phase coherence\cite{Phil03}.

On similar lines as those dealt with in a great variety of such
different approaches discussed in the literature, our aim here is
to investigate a system where the breakdown of superconductivity
situates itself in between the case of a ``bosonic'' and a
``fermionic'' mechanism for superconductivity suppression. We
focus on determining the possible ground states for interacting
Cooper pairs, in close relation to the classical notions of
superfluidity and localization of bosons and, on a more general
basis, how the phase coherence can be tuned to a phase disordered
state whose transport properties may be unconventional. In
particular, we consider within the framework of a boson-fermion
model (BFM), a system where localized bound electron pairs
(hard-core bosons) form an array of "Andreev"-like scattering
centers coupled to a fermionic subsystem of itinerant uncorrelated
electrons (quantum pair-exchange). This scenario goes beyond that
of pure phase models widely discussed in the literature, since
here one is dealing with bosonic degrees of freedom (localized
bound pairs) as well as  fermionic ones (itinerant electrons) for
which, due to the emergence  of pair correlations, one is dealing
with both amplitude as well as phase modes. The possibility to
tune from short to long-range phase coherence arises from the
following competing effects:

(i) on the one hand, the short range interaction between bosons
and fermion pairs (holes) induces a local phase locking in a
configuration with a quantum superposition between bosons and
electron pairs, leaving the common phase undetermined.

(ii) on the other hand, the itinerancy of the electrons tends to
lock and rigidly extend these initially arbitrary local phases. As
a result, phase coherence develops over longer distances by
suppressing the quantum fluctuations of the local phase, which
will involve the dynamics of amplitude fluctuations.

\noindent
Out of this competition one can recover either a superfluid state in the
regime of a small scattering rate or a phase disordered state in the
limit where the pair exchange dominates and the local quantum phase
fluctuations do not allow for long-range phase coherence. As we shall show, the
physics described by this scenario strongly depends on the concentration
of fermions and bosons, on the coupling strength, as well as on temperature
and magnetic fields. The purpose of this study is to analyze how a
transition from a superconductor to an insulator may occur and discuss
the features which can give rise to unconventional dynamics and,
eventually, unconventional transport properties in the  proximity of such
a transition.

The outline of the paper is as follows. In Sect. II we will
introduce the BFM and describe its main qualitative features and
the phenomena which can be described by it. In Sect. III, we shall
develop a path integral representation of this model and derive an
effective coarse grained action, which, after integrating out the
fermionic degrees of freedom, is able to describe the low energy
dynamics of the phase and amplitude modes. In Sect. IV we will
discuss the phase diagram based on the simplest approximation of
such a coarse grained effective action in terms of a phase-only
action and explore the transition from a superconductor to a phase
disordered state as a function of the coupling and the density of
bosons.  In Sect. V, we will study the physical features which
arise from the intrinsic Berry phase term present in our effective
action and which arises from the hard core nature of the bosons,
represented by quantum pseudo spin-$\frac{1}{2}$ variables. In
section VI we compare the salient features of this BFM scenario
with similar scenarios, such as the negative-U Hubbard model, the
Bose-Hubbard model and Josephson junction arrays. In the Discussion, section VII,
 we review the main results obtained in this paper and indicate further
developments planed for the near future.

\section{The boson-fermion model}

The boson-fermion model (BFM) in recent years has attracted
considerable attention as a model capable of capturing basic
physical properties in many body systems with strong interaction,
giving rise to the formation of resonant pair states of bosonic
nature inside a reservoir of fermions. Such a scenario was
initially proposed by one of us (JR) as an alternative to the
scenario of the hypothetical and yet to be experimentally verified
{\it bipolaronic superconductivity}. It was meant to describe the
intermediary coupling regime between the adiabatic and
anti-adiabatic limits in polaronic systems, where an exchange
between localized bipolarons and pairs of uncorrelated electrons
can be assumed to take place (for a recent intuitive justification
of such a scenario see for instance ref.\cite{Ranninger-02}).

As it has turned out, this boson-fermion scenario has a much wider
range of applicability than that for which it was initially
proposed and seems to apply to very different physical situation
such as: hole pairing in semiconductors\cite{Mysyrowicz-96},
isospin singlet pairing in nuclear matter\cite{Schnell-99}, d-wave
hole and antiferromagnetic triplet pairing in the positive-U
Hubbard model\cite{Auerbach-02} (and possibly also in the t-J
model), entangled atoms in squeezed states in molecular Bose
Einstein condensation in traps\cite{Yurovski-03} and superfluidity
in ultracold fermi gases induced by a Feshbach
resonance\cite{Timmermans-99}.

The BFM is reminiscent of an anisotropic Kondo lattice model in
terms of a pseudospin-$\frac{1}{2}$, but characterizing  localized
electron pairs instead of  localized impurity spins as in the Kondo
analogue. Its Hamiltonian is given by

\begin{eqnarray*}
H & = & (D-\mu )\sum _{i\sigma }c_{i\sigma }^{+}c_{i\sigma }
 + (\Delta _{B}-2\mu )\sum _{i}(\rho_{i}^{z}+\frac{1}{2})\nonumber\\
& + & \sum _{i\neq j,\: \sigma } t_{ij} (c_{i\sigma
}^{+}c_{j\sigma } +H.c.)+ g\sum _{i}\left(
\rho^{+}_{i}\tau^{-}_{i} + \rho^{-}_{i}\tau^{+}_{i} \right)
\end{eqnarray*}
The pseudo-spin operators $[\rho_i^+,\rho_i^-,\rho_i^z]$ denote
the local bound electron pairs (bosons) and
$[\tau_i^+=c^+_{i\uparrow}c^+_{i\downarrow},\tau_i^-=
c_{i\downarrow}c_{i\uparrow},\tau_i^z=1-\tau_i^+\tau_i^-]$ the
itinerant pairs of uncorrelated electrons.
$[c^+_{i\sigma},c_{i\sigma}]$ stand for the creation and
annihilation operators of the itinerant electrons (fermions) and
$g$ is the strength of the $boson \Leftrightarrow fermion\, pair$
exchange interaction. The hopping integral for the itinerant
electrons, which is assumed to be different from zero only for
nearest neighbor sites, is given by $t$ with a band half width
equal to $D = zt$, $z$ denoting the coordination number of the
underlying lattice. The energy level of the bound electron pairs
is denoted by $\Delta_B$. The number  of the ensemble of bosons
and fermions being conserved, $n_{tot}=n_{F\uparrow} +
n_{F\downarrow}+2 n_B$, implies a common chemical potential $\mu$
for both subsystems. $n_B$, $n_{F\uparrow,\downarrow}$ indicate
the occupation number per site of the hard core-bosons and of the
electrons with up and down spin states.

The exchange coupling between the bosons and the fermion pairs can
be considered as an effective Andreev-like scattering leading to
local states which are quantum superpositions of the form
\begin{eqnarray}
|\psi_{loc}\rangle_i =&&\int d\phi_i [\cos(\theta_i/2) \cos(\phi_i)
\rho_{i}^{+} + \sin(\phi_i)]\times \nonumber \\ && [\cos(\phi_i) +
\tau_{i}^{+}\sin(\theta_i/2)\sin(\phi_i)] |0\rangle .
\end{eqnarray}
Such states evolve gradually out of the system of localized
dephased bosons and essentially uncorrelated free fermions, which
characterize the high temperature phase of this model, when the
temperature is decreased below a certain $T^* \simeq g$ where
resonant pairing (not bound pairs!) starts to be induced in the
fermionic subsystem. These pair states have already features built
in which are reminiscent of those which characterize Cooper
pairing of fermions as well as superfluidity of bosons. The phases
of the two coherent states, corresponding to the two subsystems,
are the same and hence locked together, but are averaged over all
angles as a consequence of the conserved particle number on any
given site, $n_{tot}=2$. Roughly speaking, the ground state of the
system is then given by a product state $\prod_i
|\psi_{loc}\rangle_i$ with $cos \theta_i=1$ and which exhibits no
phase correlations on any finite length scale.  Let us next
consider the effect of fermion hopping between adjacent sites.
This will give rise to density fluctuations on each of those
individual sites and thus help to stabilize an arbitrary but
finite average value of the phases $\{\phi_i\}$ over a finite
length and time scale. In this way the localized bosons and
fermion pairs acquire
itinerancy\cite{Ranninger-95,Devillard-00,Domanski-01} which
eventually leads to a superfluid state in both
subsystems\cite{Ranninger-03}, provided that the effect of the
local correlations between the bosons and the fermion pairs can be
sufficiently diminished, but remaining still sufficiently strong
to guarantee the formation of pairing in the fermionic subsystem.
Achieving or not this situation will depend on the relative
importance of the local exchange coupling versus the fermion
hopping rate (given by the ratio $g/t$) and as well as on the
concentration of the bosons, as shown by exact diagonalization
study\cite{Cuo03} on finite clusters of this BFM.

What we shall attempt in the present study is to describe this
physics in terms of an effective action for the phase and
amplitude fields of the bosonic fields. In order to achieve this
we shall put the discussion on a level which is more familiar,
namely that one of Josephson junction arrays and Bose-Hubbard
models. For that purpose let us briefly sketch the analogy which
exists (up to a certain point) between the BFM and those systems
which have been widely discussed in the literature. A physically
possible realization of this BFM scenario can be imagined in form
of a network of superconducting grains embedded in a metallic
environment and where the only mechanism of interaction between
the grains and the fermionic background is that of Andreev
reflections. Via such a mechanism an electron (hole) is reflected
on the grain as a hole (electron) leaving behind a surplus of two
holes (electrons) in the fermionic subsystem and of two electrons
(holes) in the grain. If the grains are such that they have a
large charging energy, the fluctuations of the number of pairs on
them are energetically unfavorable and hence are largely
suppressed. We then have a situation where the state of the grains
switches essentially between zero to double occupancy with respect
to the average occupation, any time an electron (hole) is
reflected at the interface of the grain. Thus, the quantum
dynamics of the single grain can be directly represented by a
pseudospin--$\frac{1}{2}$, in order to account for the doublets
which represent the two possible states of the grains.

\begin{figure}[h]
\centerline{\psfig{figure=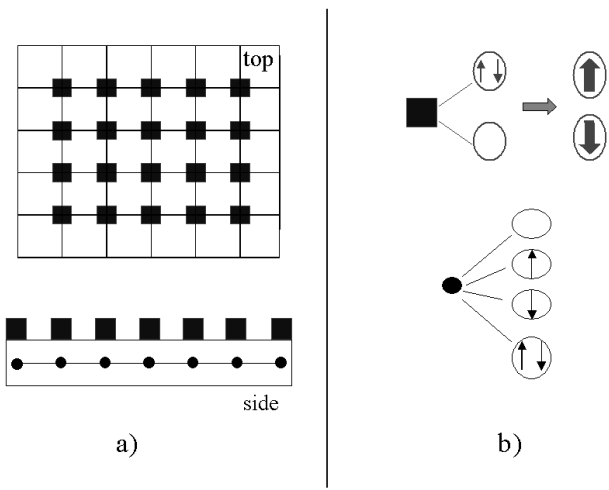 ,width=8.cm}} \caption{a)
Schematic 1D representation of the BFM on a lattice (top and side
view). The bosonic and fermionic particles  move on two different
arrays having the same periodicity:  the fermions are indicated by
circles and the bosons by squares on the respective arrays. b) The
single site configurations for the pseudospin and fermionic
variables.} \label{fig2-clustconf}
\end{figure}

For such a possible experimental setup, the effective sites in the
BFM have to be considered as defining a regular array of grains
and having the same periodicity as the underlying lattice on which
the fermions move with a hopping amplitude $t$. Moreover the size
of the grains has to be such that it is much smaller than the
distance between them. A pictorial view of such an experimental
set up in given in Fig. \ref{fig2-clustconf}. The analogy between
the BFM and the array of superconducting grains which scatter
pairs of fermions in a metallic matrix via Andreev like
reflections, may ultimately serve as an experimental device on
which to test and analyze the theoretical issues which will be
discussed in this paper.

\section{Path integral representation: derivation of the effective action}

\subsection{Generalities}

\begin{figure}[h]
\centerline{\psfig{figure=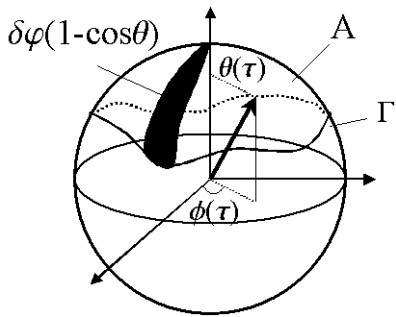 ,width=6.cm}}
\caption{Spherical representation of the pseudospin ${\bf {s}}$
including the Berry phase factor for one possible trajectory
$\Gamma$. The Berry phase term is $\exp[i s \oint (1-\cos[\theta])
\frac{\partial \phi}{\partial \tau}]=e^{i s A}$, where $A$ is the
area of the surface enclosed in the trajectory $\Gamma$. The black
part indicate the differential portion of the surface on the
sphere taken respect to the north pole.} \label{fig1-sphere}
\end{figure}

Let us now construct an effective action which describes the BFM,
with the aim to extract the dynamical properties of the low energy
degrees of freedom of the phase and amplitude modes for the bosons
and fermion-pairs. We start by expressing the partition function
in terms of a coherent-state path integral
representation\cite{Neg87}, where the fermionic part is formulated
by means of the usual Grassmann variables and the bosonic part is
described by a pseudospin-coherent state
representation\cite{Frad91,Neg87}:
\begin{eqnarray}
Z=\prod_i \int D\theta_i D\phi_i D \bar{\Psi}_i D {\Psi}_i
e^{-A[\bar{\Psi}_i,{\Psi}_i,\theta_i,\phi_i]}
\end{eqnarray}
\noindent where
\begin{eqnarray}
A[\bar{\Psi}_i,{\Psi}_i,\theta_i,\phi_i]=\int d\tau \sum_i [ i s
(1-\cos\theta_i)~\partial_{\tau} \phi_i + \nonumber\\
(\Delta_B -2 \mu )\cos\theta_i]+ \sum_{\langle ij \rangle}
\bar{\Psi}_j(\tau) G_{ij}^{-1} {\Psi}_i(\tau).
\end{eqnarray}
$\tau$ denotes the imaginary Matsubara time variable and a Nambu spinor
representation for the Grassmann
variables, related to that of the  original fermionic operators by
\begin{equation}
\begin{array}{c c c}
\bar{\Psi}_i= & \left(\begin{array}{c}
c_{i\uparrow}\\
\bar{c}_{i\downarrow}
\end{array}\right)
& {\Psi}_i= \left(\bar{c}_{i\uparrow} \, {c}_{i\downarrow} \right)
\end{array}.
\end{equation}
\noindent  The pseudospin is described by a bosonic field which in
spherical coordinates is given by ${\bf{s_i}}=s (\sin\theta_i \cos
\phi_i, \sin\theta_i \sin \phi_i, \cos\theta_i)$ (see Fig.
\ref{fig1-sphere}). $\theta_i$ describes the polar angle of the
vector ${\bf s}_{i}$ with respect to the north pole of the $z$
axis, while $\phi_i$ is the azimuthal angle which defines the
angular position of the basal plane projection of this vector. The
first term of the action $A$ is the Wess-Zumino
term\cite{Frad91,Stone86}, ensuring the correct quantization of
the quantum pseudospin variable. For any path, parameterized by
$\phi(\tau)$ and $\theta(\tau)$, the contribution of this term is
equal to $i\,s$ times the surface area of the sphere between this
path and the north pole. For closed paths this has exactly the
form of the Berry phase \cite{Hal86}.  The second term is linked
to the density of the bosons $n_B(\tau)$ through the $\cos
\theta(\tau)$ dependence of the pseudospin. Finally, the last
contribution of the action contains the coupling between the
fermionic and bosonic subsystem through the Green's function
$G_{ij}$, determined by:
\begin{equation}
\begin{array}{c c}
\left(\begin{array}{c c}
K_1 & L\\
L^{*} & K_2
\end{array} \right) & G_{ij}(\tau-\tau')=\delta(\tau-\tau')
\end{array}
\end{equation}
\noindent where $K_1=(-\partial_{\tau}+\mu)\delta_{ij}+ t_{ij}$,
$K_2=(-\partial_{\tau}-\mu)\delta_{ij}- t_{ij}$ and $L=g \,
\sin\theta_i(\tau) e^{i\,\phi_i(\tau)} \delta_{ij}$ and  $L^{*}$ being
the conjugate field of $L$. Integrating out the fermionic part,
one obtains the action in terms of exclusively the bosonic fields:
\begin{eqnarray}
\nonumber A=&&-Tr \ln G^{-1}+\int d\tau \sum_i [ i s
(1-\cos\theta_i(\tau))~
\partial_{\tau} \phi_i(\tau) +\\
&&(\Delta_B -2 \mu )\cos\theta_i(\tau)]
\end{eqnarray}
where the trace has to be carried out over all internal as well as
space-time indices.

%
\begin{figure}[h]
\centerline{\psfig{figure=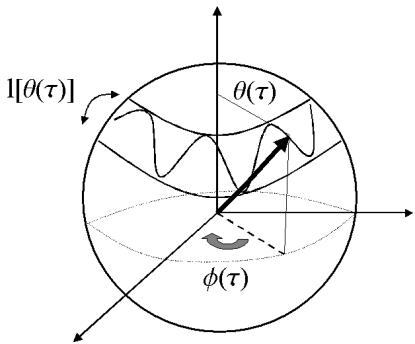 ,width=5.5cm}} \caption{
Representation of a possible path for the local pseudospin motion.
The field $l[\theta(\tau)]$ indicates the undulation of the
pseudospin vector along the polar direction as it arises from the
fluctuations of the average boson density ($\langle \cos \theta
\rangle$) and of the pairing amplitude ($\langle \sin \theta
\rangle$), while it precesses around the $z$-axis due to the time
evolution of the phase variable $\phi(\tau))$.}
\label{fig2-sphere}
\end{figure}
%

Up to this point no approximation has been made in the derivation
of the action which describes the coupling between the bosonic and
fermionic degrees of freedom. It is important to notice that the
variation of the bosonic variable $\theta(\tau)$ describes both,
the density fluctuations of the bosonic subsystem (via the
projection of the pseudospin vector on the $z$ axis, i.e.
$\cos\theta(\tau)$, being the longitudinal component)  and the
amplitude fluctuations of the fermionic pair field (via the
transverse part as projection of the pseudospin vector onto the
basal plane, i.e. $\sin\theta(\tau)$) (see Fig.
\ref{fig1-sphere})). The variable $\phi(\tau)$ determines the
rotational degrees of freedom of the pseudospin vector, expressing
its phase dynamics.

For extracting the relevant terms which control the low energy
dynamics of the coupled phase and amplitude modes and performing
an expansion, which is meaningful in terms of the phase variable,
it is judicious to make the following steps: i) gauge away the
phase dependence from the term $L$ which permits to separate the
trace into a part which does not depend on the phase of the
bosonic field and another part that contains only spatial and time
variations of $\phi_i(\tau)$\cite{r1}, ii) rewrite the term $L$
after the gauge transformation, as a sum of two pieces, one not
dependent on time (which is linked to the average density of
bosons) and  another term containing the fluctuations with respect to its
mean value.

The first operation is performed by applying to the operators
under the trace the rotation $U_i=e^{i\,\phi_i(\tau) \sigma_3/2}$,
where $\sigma_i$ denote the Pauli matrices.
Hence,
\begin{eqnarray}
Tr \ln G^{-1}=Tr \ln U \tilde{G}^{-1} U^{-1}=Tr \ln \tilde{G}^{-1}
\end{eqnarray}
where
\begin{eqnarray}
\nonumber \tilde{G}_{ij}^{-1}&=&\left[-\partial_{\tau} \sigma_0
+[\frac{i}{2}
\partial_{\tau}\phi_i(\tau) - \mu] \sigma_3+g~\sin\theta_i(\tau)
~\sigma_1\right]\delta_{ij} +\\&& t_{ij}
e^{i\,[\phi_i(\tau)-\phi_j(\tau)] \sigma_3/2}.
\end{eqnarray}

Since the bosonic density is fixed in average, we now separate the part
which depends on the polar angle in a
time independent contribution and its time dependent correction. That is, the
term $\sin\theta_i(\tau)$ is decomposed into its average value
$\langle \sin\theta_{i} \rangle$ (which is determined by fixing the
density of bosons due to the spherical constraint) plus a time
dependent contribution $l[\theta(\tau)]$ which contains the
fluctuation around its average value (Fig. \ref{fig2-sphere}). Thus
this local field, due to the constraint, will
describe both: the time dependent variation of the density as well as  of
the pairing amplitude.

Next, let us write the Green's function in the usual form as
\begin{eqnarray}
\tilde{G}_{ij}^{-1}&=& {G_0}_{ij}^{-1}+\Sigma_{ij}
\end{eqnarray}
\noindent with ${G_0}_{ij}^{-1}=\left[-\partial_{\tau} \sigma_0 - \mu \sigma_3
+\tilde{g} ~\sigma_1\right]\delta_{ij}$ and
$\Sigma_{ij}=T_{ij}+D_i+K_i$, where
\begin{eqnarray}
D_i&=&\frac{i}{2} \frac{\partial \phi(\tau)}{\partial \tau}
\sigma_3\\ K_i&=& \tilde{g} ~l[\theta(\tau)] \sigma_1\\
T_{ij}&=&t_{ij} e^{ -i (\phi_i(\tau)-\phi_j(\tau))
\frac{\sigma_3}{2}}\\ \tilde{g}&=&g \langle\sin[\theta]\rangle.
\end{eqnarray}

From  this point onward we shall assume the average density of
bosons to be homogeneously distributed and thus given by
$n_B=\frac{1}{2}(1+\langle\cos\theta \rangle)$. This implies that
the bare coupling $g$ is renormalized to $\widetilde g$ as a
consequence of the spherical constraint of the pseudospin
variable. Moreover, since $\langle \sin\theta \rangle =
\sqrt{1-\langle\cos \theta\rangle ^2}$ and the density of bosons
is fixed in average via a suitable choice of the chemical
potential and of the on-site bosonic energy, one can treat the
exchange coupling as an external boson-density tunable parameter.
The variation in the $\theta$ variable is then simply related to
the variation of the bosonic density such that if $\theta$ varies
in the range $[0,\pi]$ then $n_B$ varies in the interval $[0,1]$.

Before expanding the trace, let us write down explicitly the
expression of the zero order Green's function, as it will be frequently
used in the following steps:
\begin{eqnarray}
{G_0}_i^{\alpha\beta}(\tau)=\frac{1}{\beta} \sum_{\omega_n}
{G_0}_i^{\alpha \beta}(\omega_n)~\exp[-i~\omega_n~\tau]
\end{eqnarray}
\noindent with
\begin{equation}
\begin{array}{c c}
{G_0}_i^{\alpha \beta}(\omega_n)= & \left(\begin{array}{c c}
\frac{-i \omega_n+\mu}{\omega_n^2+\omega_0^2} &
\frac{\sqrt{\omega_0^2-\mu^2}}{\omega_n^2+\omega_0^2}\\
\frac{\sqrt{\omega_0^2-\mu^2}}{\omega_n^2+\omega_0^2} & \frac{-i
\omega_n-\mu}{\omega_n^2+\omega_0^2}
\end{array} \right)
\end{array}
\end{equation}
and where  we introduced  $\omega_0 = \tilde{g}$.
\subsection{Second order loop expansion}
 We now evaluate the contribution of the self energy $\Sigma_{ij}$ to the
effective action. This is done in the usual way by making a loop
expansion in the trace\cite{r1}. We shall construct that expansion
up to second order in the time and space derivatives of the phase
variable and in the terms which contain both, the fluctuations of
the density and the amplitude. For that purpose we use the
standard identity:
\begin{eqnarray}
\nonumber Tr \ln \tilde{G}^{-1}&=&Tr \ln \left[
{G_0}^{-1}+\Sigma\right]\\
&=&Tr \ln {G_0}^{-1}+Tr \ln \left[ 1+ {G_0}
\Sigma\right]
\end{eqnarray}
and then expand the second term of this expression
up to second order in $\Sigma$, such as to keep all the
contributions up to quadratic order in the gradient of the phase. This gives
\begin{eqnarray}
Tr \ln \tilde{G}^{-1}\cong Tr \ln {G_0}^{-1}+Tr [{G_0} \Sigma]
-\frac{1}{2} Tr [{G_0} \Sigma]^2 \,.
\end{eqnarray}
The first term of this expression is just a constant and does not
contribute to the dynamics. In the second term
\begin{eqnarray}
Tr [{G_0} \Sigma]=Tr [{G_0} T]+Tr [{G_0} D]+Tr [{G_0} K]
\end{eqnarray}
the only parts different from zero are $Tr [{G_0}_iK_i]$ and
$Tr [{G_0}_i D_i]$.  $Tr [{G_0}_i T_{ij}]$ gives no contribution once
one makes the trace over the site indices.
$Tr [{G_0}_i D_i]$  introduces a contribution which is proportional
to the chemical potential multiplied by the time derivative of the phase
$\sim i \mu \partial_{\tau} \phi$. We will see below that this contribution
describes an effective {\it off-set charge} (in terms of the terminology of
a similar Josephson junction array scenario) and which arises from the
total fermionic and bosonic static
density distribution via their dependence on the chemical potential.
$Tr [{G_0}_i K_i]$ describes the lowest order fluctuations of
the bosonic (fermion pair) density  due to the presence of the
spontaneous pair/hole creation out of the condensate. Its direct evaluation
gives a contribution
\begin{eqnarray}
A_1=\int d\tau_1 d\tau_2 Tr_{ps}[{G_0}_i(\tau_1,\tau_2)
K_i(\tau_2,\tau_1)]
\end{eqnarray}
where $Tr_{ps}$ represents the trace over the internal pseudospin index.
With $K_i(\tau_2,\tau_1)=K_i(\tau_1) \delta(\tau_1-\tau_2)$ and
integrating over the Matsubara times gives:
\begin{eqnarray}
\nonumber A_1&&=\int d\tau Tr_{ps}[{G_0}_i(0) K_i(\tau_)] \\ \nonumber
&&=2 g {G_0}_i^{12}(0) \int d\tau l[\theta(\tau)]\\&&= -\frac{g
\tanh[\beta \omega_0]}{\omega_0} \int d\tau l[\theta(\tau)],
\end{eqnarray}
or in a compact form
\begin{eqnarray}
A_1&&= E_1 \int d\tau l[\theta(\tau)] \nonumber \\
E_1&&=-\frac{g \tanh[\beta \omega_0]}{\omega_0}.
\end{eqnarray}

Let us next come to the evaluation of  the terms which contribute to the
quadratic order in this loop expansion of the trace. The
parts which are non zero in these terms are the following:
\begin{eqnarray}
A_2&=&Tr~[{G_0}_i~K_i~{G_0}_i~D_i] \nonumber \\
A_3&=&Tr~[{G_0}_i~K_i~{G_0}_i~K_i] \nonumber \\
A_4&=&Tr~[{G_0}_i~D_i~{G_0}_i~D_i] \nonumber \\
A_5&=&Tr~[{G_0}_i~T_{ij}~{G_0}_j~T_{ji}]
\end{eqnarray}

The term $A_2$  contains processes where
the fermionic background locally couples at different times
to the fluctuations of the bosonic density and to the phase fluctuations:
\begin{eqnarray}
\nonumber &&A_2=Tr~[G_{0\,i}~K_i~G_{0\,i}~D_i] =\qquad \\
&&\int d\tau_1d\tau_2 Tr_{ps}[G_{0\,i}(\tau_1-\tau_2)
K_i(\tau_2) G_{0\,i}(\tau_2-\tau_1)D_i(\tau_1)]. \nonumber \\
\end{eqnarray}
Since ${G_0}_i(\tau_1-\tau_2)$ depends exclusively on the
time differences, we introduce the new variables
$\tau=\frac{\tau_1-\tau_2}{2}$ and $\eta=\frac{\tau_1+\tau_2}{2}$,
after which the integral over the Matsubara time variables becomes:
\begin{equation}
\nonumber A_2=\frac{1}{2} \int d\tau d\eta Tr_{ps}[{G_0}_i(\tau)
K_i(\eta-\tau) {G_0}_i(-\tau) D_i(\tau+\eta)].
\end{equation}
As we are interested in the gradient expansion in the bosonic
phase and density, related to  $\phi(\tau)$ and $\theta(\tau)$, we
keep only the lowest order in their time derivatives. This is done
by considering the expansion of the product
$K_i(\eta-\tau)~D_i(\tau-\eta)$ around $\tau=0$ in order to
separate the parts which depend exclusively on the local variable
of the Green function and those which are linked to the phase and
the density fluctuations. The expansion then reads as follows,
\begin{eqnarray}
K_i(\eta-\tau)~D_i(\eta+\tau)& \simeq &
K_i(\eta)~D_i(\eta)+\tau \{-\frac{\partial K_i(\eta)}{\partial
\eta} D_i(\eta) \nonumber \\
&&+\frac{\partial D_i(\eta)}{\partial \eta}
K_i(\eta)\}+O(\tau^2)
\end{eqnarray}
Due to the symmetry of the Green's functions (even
for $\tau \rightarrow -\tau$), the linear contribution of
the series expansion of $K_i(\eta-\tau)~D_i(\tau-\eta)$ cancels in
the effective action after integrating over the time. Hence, to
lowest order one obtains,
\begin{eqnarray}
A_2& = & \frac{1}{2} \int d\tau d\eta Tr_{ps}[{G_0}_i(\tau) K_i(\eta)
{G_0}_i(-\tau) D_i(\eta)] \nonumber \\
& = & \frac{g}{2} \int d\tau \left[{G_0}_i^{12}(\tau)
{G_0}_i^{11}(-\tau) + {G_0}_i^{11}(\tau) {G_0}_i^{12}(-\tau) \right.\nonumber \\
&-& \left. {G_0}_i^{22}(\tau) {G_0}_i^{12}(-\tau) -
{G_0}_i^{12}(\tau) {G_0}_i^{22}(-\tau)\right]\times \nonumber \\
& &\int
d\eta \frac{i}{2} \frac{\partial \phi(\eta)}{\partial \eta}
l[\theta(\eta)]
\end{eqnarray}
\noindent or in a compact form,
\begin{equation}
A_2 = E_2 \int d\eta ~i~ \frac{\partial \phi(\eta)}{\partial
\eta}~ l[\theta(\eta)]
\end{equation}
\noindent where the coefficient $E_2$, after integrating over the
Matsubara time, is given by:
\begin{equation}
E_2=\frac{\tilde{g}^2}{4}
\frac{\mu~(\beta~\omega_0-\sinh{[\beta~\omega_0]})}
{\omega_0^3(1+\cosh{[\beta~\omega_0])}}.
\end{equation}

Let us next consider the term $A_3$, which expresses the
coupling of the fermionic field to the fluctuations of bosonic density at
different times. In order to extract the lowest order gradient
contributions, we follow the same procedure as that just used above, giving us:
\begin{equation}
A_3=\frac{1}{2} \int d\tau~d\eta Tr_{ps}[{G_0}_i(\tau) K_i(\eta-\tau)
{G_0}_i(-\tau) K_i(\tau+\eta)]
\end{equation}
Performing again the expansion in time up to quadratic order
in the time derivatives of $K_i(\tau)$, we have:
\begin{equation}
K_i(\eta-\tau)~K_i(\eta+\tau)\simeq K_i(\eta)^2 -\tau^2
\frac{\partial K_i(\eta)}{\partial \eta}^2.
\end{equation}
and hence can rewrite $A_3$ in the following way:
\begin{eqnarray}
&& A_3=\frac{1}{2} \int d\tau~d\eta Tr_{ps}[{G_0}_i(\tau) K_i(\eta)
{G_0}_i(-\tau) K_i(\eta)]+ \nonumber \\
&&\frac{1}{2} \int d\tau~d\eta~
Tr_{ps}[{G_0}_i(\tau) K^{'}_i(\eta) {G_0}_i(-\tau) K_i^{'}(\eta)].
\end{eqnarray}
Carrying out the trace over the internal indices gives:
\begin{eqnarray}
Tr_{ps}[{G_0}_i(\tau) K_i(\eta) {G_0}_i(-\tau)
K_i(\eta)]=[{G_0}_i^{22}(\tau) {G_0}_i^{11}(-\tau)  \nonumber \\
 + 2~{G_0}_i^{12}(\tau) {G_0}_i^{12}(-\tau)+{G_0}_i^{11}(\tau)
{G_0}_i^{22}(-\tau)]~(\tilde{g}^2 l[\theta(\eta)]^2 ) \nonumber \\
\end{eqnarray}
so that
\begin{equation}
A_3=E_{3a} \int ~d\eta~ l[\theta(\eta)]^2- E_{3b} \int ~d\eta
\frac{\partial ~l[\theta(\eta)]}{\partial \eta}^2.
\end{equation}
\noindent By evaluating the integrals over the Matsubara times we obtain:
\begin{eqnarray}
E_{3a}&&=\frac{\tilde{g}^2}{2} \frac{\beta \omega_0
\tilde{g}^2+\mu^2 \sinh[\beta \omega_0]}{\omega_0^3
(1+\cosh[\beta~\omega_0])} \nonumber \\
E_{3b}&&=\frac{\tilde{g}^2}{2}[ \frac{sech[(\beta
\omega_0/2)]^2 ( -6 \beta \mu^2 \omega_0+\beta^3 ( -\mu^2
\omega_0^3 +\omega_0^5)}{24 \omega_0^5} \nonumber \\
 &&+\frac{6\mu^2 \sinh[\beta \omega_0])}{24 \omega_0^5}].
\end{eqnarray}

Applying the same procedure for the evaluation of $A_4$ (which has
the same functional form as $A_3$ but involving the coupling
between the fermionic degrees of freedom and the phase velocity at
different time) one gets an analogous expression, providing we
discard terms of higher order in the derivatives than
$[\frac{\partial\phi(\tau)}{\partial \tau}]^2$:
\begin{eqnarray}
A_4&& \simeq \frac{1}{2} \int d\tau~d\eta
Tr_{ps}[{G_0}_i(\tau) D_i(\eta) {G_0}_(-\tau) D_i(\eta)] \nonumber \\
&&=\frac{1}{8} C_0 \int ~d\eta (\frac{\partial
\phi(\eta)}{\partial \eta})^2.
\end{eqnarray}
\noindent Here
\begin{eqnarray}
 C_0=&& \int d\tau \{-[{G_0}_i^{11}(\tau) {G_0}_i^{11}(-\tau)-
2~{G_0}_i^{12}(\tau) {G_0}_i^{12}(-\tau)+\nonumber \\
&&  {G_0}_i^{22}(\tau){G_0}_i^{22}(-\tau)] \}~,
\end{eqnarray}
which after evaluating the integration over the Matsubara times gives,
\begin{equation}
C_0=\frac{sech[(\beta \omega_0/2)]^2 (\beta \mu^2
\omega_0+\tilde{g}^2 \sinh[\beta \omega_0])}{2 \omega_0^3}.
\label{selfcapac}
\end{equation}

Finally, we come to the evaluation of the last term $A_5$  which
involves the coupling between phase fluctuations on different
sites and the process of single particle hopping. This
contribution will yield terms which are quadratic in the time
derivative of the phase ({\it charging like}, in the terminology of a similar
Josepson junction array scenario) and moreover will
generate an effective hopping induced inter-site phase coupling
which turns out to be similar to the Josepshon coupling in arrays
of superconducting grains.

Considering again the lowest order gradient expansion contributions we have:

\begin{eqnarray}
\nonumber A_5&&=Tr~[{G_0}_i~T_{ij}~{G_0}_j~T_{ji}]\\\nonumber &&\simeq
\frac{1}{2} \int d\tau~d\eta Tr_{ps} \left\{ {G_0}_i(\tau)
T_{ij}(\eta) {G_0}_j(-\tau) T_{ji}(\eta)-\right.\\ \nonumber && \left.
\tau^2 \left[{G_0}_i(\tau) \frac{\partial T_{ij}(\eta)}{\partial \eta}
{G_0}_j(-\tau) \frac{\partial T_{ij}(\eta)}{\partial \eta}\right]\right\}\\
&&=P_1+P_2
\end{eqnarray}
The terms $P_{1,2}$ are given by:
\begin{eqnarray}
&P_1&=-{t^2_{ij}} \int d\tau~[{G_0}_i^{12}(\tau)
{G_0}_i^{12}(-\tau)]~\int d\eta \cos[\phi_i(\eta)-\phi_j(\eta)] \nonumber\\
&P_2&=-\frac{t^2_{ij}}{8} \int d\tau~
\tau^2~[G_{0\,i}^{11}(\tau) G_{0\,j}^{11}(-\tau)+G_{0\,i}^{22}(\tau)
G_{0\,j}^{22}(-\tau)]~\nonumber\\
&& \int d\eta [\frac{\partial
\phi_i(\eta)}{\partial \eta}-\frac{\partial \phi_j(\eta)}{\partial
\eta}]^2.
\end{eqnarray}
By carrying out the integration over the Matsubara times one
obtains for the hopping induced inter-site amplitude and phase
coupling:
\begin{eqnarray}
&P_1&=E_{J} \int d\eta \cos[\phi_i(\eta)-\phi_j(\eta)] \nonumber \\
&E_{J}&=\frac{t^2_{ij}~\tilde{g}^2 \left( -\beta \omega_0
sech[\beta \omega_0/2]^2 +2 \tanh[\beta \omega_0/2] \right)}{8
\omega_0^3}.\nonumber \\
\label{joseph}
\end{eqnarray}

Similarly, by calculating the coefficient of the term $P_2$, one obtains
the strength of a {\it mutual capacitance} term (in the terminology  of a
similar Josephson junction array scenario) for neighboring effective sites:
\begin{eqnarray}
\nonumber P_2&&=\frac{1}{8} C_{1}~\int d\eta [\frac{\partial
\phi_i(\eta)}{\partial
\eta}-\frac{\partial \phi_j(\eta)}{\partial \eta}]^2 \\
\nonumber C_{1}&&=-t^2_{ij} \int_{-\beta/2}^{\beta/2} d\tau~
\tau^2~[G_i^{11}(\tau) G_i^{11}(-\tau)+G_i^{22}(\tau)
G_i^{22}(-\tau)]\\
\nonumber &&=-{t^2_{ij}} \left[ \frac{\exp[\beta
\omega_0] (-6 \beta \tilde{g}^2 \omega_0+ \beta^3 \omega_0^3
(\mu^2+\omega_0^2)}{12 (1+\exp[\beta \omega_0])^2
\omega_0^5} \right.\\
 &&+\left. \frac{6 \tilde{g}^2 \sinh[\beta
\omega_0])}{12 (1+\exp[\beta \omega_0])^2 \omega_0^5} \right].
\label{mutualcap}
\end{eqnarray}

The final effective action is then given by the sum of all the
terms  evaluated above which are grouped together in form of
three different contributions:
\begin{eqnarray}
S=\int d\tau \left[S_{\phi}+S_{\phi-\theta}+S_{\theta} \right]. \,
\label{action}
\end{eqnarray}
$S_{\phi}$, $S_{\theta}$, $S_{\phi-\theta}$ are the contributions
arising from exclusively (i) the phase dynamics, (ii) the
fluctuations of the bosonic density and of the amplitude
$l[\theta]$ and (iii) the coupling between them. They are given
by:
\begin{eqnarray}
S_{\phi}&=&\sum_{i} \frac{1}{8}(C_0+z C_{1})~(\frac{\partial
\phi_i(\tau)}{\partial \tau})^2 \nonumber \\
&-&\sum_{\langle i,j\rangle} \left[\frac{1}{8}C_{1} \frac{\partial
\phi_i(\tau)}{\partial \tau} \frac{\partial \phi_j(\tau)}{\partial
\tau} + E_{J}~ \cos[\phi_i(\tau)-\phi_j(\tau)]\right] \nonumber \\
&+& i \sum_{i}\mu \frac{\partial
\phi_(\tau)}{\partial \tau}\nonumber \\
S_{\theta}&=&\sum_i \left[E_1~l[\theta_i(\tau)] +E_{3a}
l[\theta_i(\tau)]^2-E_{3b} [\frac{\partial
~l[\theta_i(\tau)]}{\partial \tau}]^2\right] \nonumber \\ \nonumber
S_{\phi-\theta}&=&\sum_i\left[ i~s\,\frac{\partial
\phi_{i}(\tau)}{\partial \tau}~(1-\cos[\theta_i(\tau)]) \right. \\
&+& \left.i~E_2\frac{\partial \phi_i(\tau)}{\partial \tau}~ l[\theta_i(\tau)]
\right].
\label{action1}
\end{eqnarray}
We expand the Berry phase contribution (the first term in
$S_{\phi-\theta}$)  up to second order in
$l[\theta(\tau)]$  and subsequently redefine this field as
${l}[\theta(\tau)] = a + b \bar{l}[\theta(\tau)] $, with the time independent
constants $a,b$ chosen in  such a way as to eliminate any terms linear in
$\bar{l}$ in the the action $S$. We then find two contributions to the action
which are linear in ${\partial \phi(\tau)}/{\partial \tau}$: One which is time
independent and which can be absorbed into the  chemical potential and
another one which is quadratic in $\bar{l}[\theta(\tau)]$.
With this, $S_{\phi-\theta}$ can be rewritten as:
\begin{eqnarray}
S_{\phi-\theta}&=&\sum_i i~\,\frac{\partial
\phi_{i}(\tau)}{\partial \tau} \bar{q}_i(\tau) \, ,
\label{action2}
\end{eqnarray}
where  $\bar{q}_i(\tau)$ is quadratic in $\bar{l}[\theta(\tau)]$.

The effective action thus constructed for the BFM (Eqs.
(\ref{action},\ref{action1})) is similar to that of Josephson
junction arrays with nearest neighbor, as well as, self {\it
capacitance}, {\it Josephson} coupling and  {\it off-set charge}
terms. The action for the BFM goes however beyond that for such
Josephson junction arrays in the following respects. We have extra
terms which control the dynamics of the amplitude modes, given by
$S_{\theta}$ and an intrinsic Berry phase term which gives rise to
a direct phase - amplitude coupling (Eq. \ref{action2}), where the
dynamical amplitude fluctuations would correspond to a time
dependent {\it offset charge} term in an analogous Josephson junction
array picture. Finally, the Berry phase term, being an intrinsic
topological term, may give rise to a Magnus force on a vortex, as
it will be discussed in section V.

%


\section{THE SUPERCONDUCTOR - INSULATOR PHASE BOUNDARY}
In a first approach  to analyze the stability region of long-range
phase superconducting coherence, we examine the  effective action
(eqs. \ref{action},\ref{action1}), restricting ourselves to the
phase only part of it. This means a study of the SIT  driven by a
competition between the phase coherence induced by pair hopping
and the disrupting effect of local boson density (or equivalently
pair field amplitude) fluctuations in the presence of a source
term for the bosons which controls the global boson density via an
effective chemical potential. Within such an approximation our
study is equivalent to that of Josephson junction arrays,
\underline{except} that the effective coupling constants entering
in such an action depend in a highly non trivial way on the
parameters which characterize the original BFM Hamiltonian. This,
as we shall see, will lead to novel features concerning the phase
diagram with a SIT for the BFM when we examine it in terms of the
boson-fermion exchange coupling $g/t$ and the boson concentration
$n_B$.

We shall determine the phase diagram by means of the so-called
coarse graining approximation which has been successfully applied
for this kind of problem and which permits to capture the relevant
qualitative and quantitative features of such a Josephson junction
array like action\cite{Don81,Brud92}. It is known  that, as a
consequence of the uncertainty relation between the phase $\phi_i$
and the pair number operator $Q_i=i \frac{\partial}{\partial
\phi_i}$, the system can switch from a phase ordered  to a
disordered state. An essential part of this study will concern how
the relevant parameters of the BFM Hamiltonian influence the
equivalent amplitudes of the Josephson coupling, the {\it
capacitance} and the {\it off-set charge} terms. Thus, at zero
temperature, by fixing the bosonic density distribution, we find
that the variation of the ratio between the Josephson coupling
energy  $E_J$ and the {\it charging} energy $E_C=C^{-1}_0/2$,
which controls the phase-density interplay, increases from zero,
goes through a maximum and then decreases to zero with increasing
$g/t$ . Or else, if one fixes the coupling $g/t$, then by varying
the density of the bosonic distribution one is able to control the
effective coupling which appears in $E_J$ and $C_{ij}$  bt varying
the average bosonic density $\tilde{g}=2 g \sqrt{n_B (n_B-1)}$).
It is thus immediately evident that in the BFM scenario there is a
non-trivial interplay between the renormalization of the Josephson
coupling and the charging effect. If one goes to the limit of
empty ($n_B=0$) and full bosonic occupation ($n_B=1$), the
effective coupling $\tilde{g} \rightarrow0$, and the critical
temperature consequently reduces to zero.

The general form of the action we then have to examine is given by:
\begin{eqnarray}
\nonumber S_{phase}=&&\int_{0}^{\beta} \sum_{i} \frac{i}{2}
\frac{\partial \phi_i(\tau)}{\partial \tau} q_i -
\\
\nonumber &&
E_{J}\sum_{ij}\alpha_{ij} \cos[\phi_i(\tau)-\phi_j(\tau)]+
\\ &&
\sum_{i,j}\frac{1}{8}\frac{\partial \phi_i(\tau)}{\partial \tau} C_{ij}
\frac{\partial \phi_j(\tau)}{\partial \tau}.
\end{eqnarray}
The first term describes the effect of a static {\it off-set
charge} ($q_i$) The second term contains the physics of the pair
hopping processes, an analog to the Josephson tunnelling
processes, and has a coupling strength $E_J$ with $\alpha_{ij}=1$
if $(i,j)$ are nearest neighbors and zero otherwise. Finally, the
third term describes the {\it charging} term arising from the
local exchange of boson and fermion pairs and from  quasi-particle
hopping between nearest neighbor sites. The strength of this
{\it charging} type interaction is given by  $C_{ij}=(C_0+z C_1)
\delta_{i,j}-C_1 \sum_p \delta_{i,i+p}$ which represents an
effective general {\it capacitance} matrix, with the vector $p$ running
over the nearest neighbors. $C_0$ denotes the {\it self-capacitance} and
$C_1$ the mutual one ($z$ being the coordination number).

As mentioned before, to extract the phase diagram we make use of
the coarse graining approximation\cite{Don81}. The main idea of
this approach is to introduce a Hubbard-Stratonovich auxiliary
field which is conjugate to the average of $\langle e^{i \phi_i}
\rangle$ and which plays the role of an order parameter for the
transition from a superconducting to an insulating state. Since
the phase transition has a continuous character, one can expand
the action in powers of the auxiliary field and determine the
occurrence of phase coherence by looking at the coefficients of
the quadratic-term in the limit of long-wavelengths and zero
frequency.

We briefly sketch the main steps of such an  approximation and adapt it to
the present scenario of the BFM. The partition
function for $S_{phase}$ is given by the sum of all the possible
paths of the phase variables in the imaginary time and in the real
space:
\begin{eqnarray}
Z=&&\int \prod_i D\phi_i \exp\left\{ \int_{0}^{\beta}
\sum_{i} -\frac{i}{2} \frac{\partial \phi_i(\tau)}{\partial \tau} q_i(\tau)
\nonumber\right.\\
&& +\left.E_{J} \sum_{ij}\alpha_{ij}
\cos[\phi_i(\tau)-\phi_j(\tau)]\right.\nonumber\\
\nonumber && -\left.\frac{1}{8} \sum_{ij} \frac{\partial \phi_i(\tau)}{\partial
\tau} C_{ij} \frac{\partial \phi_j(\tau)}{\partial \tau} \right\}
\end{eqnarray}

To perform the Hubbard-Stratonovich transformation, one rewrites
the {\it Josephson coupling} term as $\frac{E_J}{2}\sum_{ij}
\exp[i \phi_i] \alpha_{ij} \exp[i \phi_j]$ and then, by using the
usual Gaussian identity, introduces an auxiliary field
$\psi_i(\tau)$. The partition function $Z$ then becomes:
\begin{eqnarray}
&&Z=\int \prod_i D\psi_i^{*} D\psi_i D\phi_i
\exp\left\{\int_{0}^{\beta} d\tau [ (\frac{-2}{E_J}) \sum_{ij} \psi_i^{*}
\alpha^{-1}_{ij} \psi_j\right\}  %
\nonumber \\ \nonumber
&& \exp\left\{ \int_{0}^{\beta}  d\tau \left[
\sum_{i} \frac{i}{2} \frac{\partial \phi_i(\tau)}{\partial \tau}
q_i(\tau)
-\sum_i (\psi_i e^{i \phi_i}-\psi^{*} e^{-i \phi_i}) \right.\right.\\
 && -\left. \left. \sum_{i,j}\frac{1}{8} \sum_{ij} \frac{\partial \phi_i(\tau)}
{\partial
\tau} C_{ij} \frac{\partial \phi_j(\tau)}{\partial \tau} \right]\right\}.
\end{eqnarray}

Hence, starting from the effective action for the auxiliary field,
one can perform an expansion up to second order in $\psi$ and thus
derive the usual Ginzburg Landau type free energy functional which permits
to determine the boundary line between the superconducting and the insulating
state. The corresponding effective action for these auxiliary fields
$\psi, \psi^*$
\begin{eqnarray}
S_{\psi}=&&\ln \left[ \int \prod_i D\phi_i
\exp \left( \int_{0}^{\beta} d\tau \left\{
\sum_{i} \frac{i}{2} \frac{\partial \phi_i(\tau)}{\partial \tau}
q_i(\tau) \right. \right. \right. \nonumber\\
&-& \left. \left. \left. \sum_i (\psi_i(\tau) e^{i
\phi_i(\tau)}-\psi_i^{*}(\tau) e^{-i \phi_i(\tau)}) \right. \right. \right. \nonumber\\
&-& \left. \left. \left. \sum_{i,j}\frac{1}{8}\frac{\partial \phi_i(\tau)}{\partial
\tau} C_{ij}
 \frac{\partial \phi_j(\tau)}{\partial \tau} \right\} \right ) \right]
\end{eqnarray}
is then expanded to second order, giving:
\begin{eqnarray}
S_{\psi}=\int_{0}^{\beta} d\tau d\tau^{'} \chi_{ij}(\tau,\tau^{'})
\psi_i^{*}(\tau) \psi_j(\tau)+O(\psi^4),
\end{eqnarray}
\noindent where
\begin{equation}
\chi_{ij}(\tau,\tau^{'})=\langle e^{i \left[
\phi_i(\tau)-\phi_j(\tau)^{'} \right]} \rangle_0
\end{equation}
denotes the two-time phase correlator,
which is equal to the second derivative with respect to the
auxiliary field $\psi$ and its conjugate at different time and
space positions, evaluated around their zero values. By performing
the functional derivation, one gets the following expression for it:

\begin{eqnarray}
\chi_{ij}(\tau,\tau^{'})=\frac{\int \prod_i D\phi_i e^{i \left[
\phi_i(\tau)-\phi_j(\tau)^{'} \right]} \exp[-S_0]}{\int \prod_i
D\phi_i \exp[-S_0]}
\end{eqnarray}
with $S_0$ being the part of the action which contains exclusively the
{\it charging} contributions, i.e. :
\begin{eqnarray}
S_0&=&\exp\left[ \int_{0}^{\beta} d\tau \sum_{i} \frac{i}{2}
\frac{\partial \phi_i(\tau)}{\partial \tau} q_i(\tau) \right. \\
\nonumber & & -\left. \sum_{i,j}\frac{1}{8}\frac{\partial
\phi_i(\tau)}{\partial \tau} C_{ij} \frac{\partial
\phi_j(\tau)}{\partial \tau} \right].
\end{eqnarray}
According to the scheme outlined above, we now develop the
partition function $Z$ up to second order in the auxiliary fields,
thus putting it into a familiar form:
\begin{eqnarray}
Z=\int \prod_i D\psi_i^{*} D\psi_i e^{-F_\psi}
\end{eqnarray}
with
\begin{equation}
F_\psi=\int_{0}^{\beta}  d\tau  d\tau^{'} \sum_{ij}
\psi_i^{*}(\tau) \left[ \alpha_{ij} \delta(\tau-\tau^{'})-
\chi_{ij}(\tau,\tau^{'}) \right] \psi_i(\tau^{'}).
\end{equation}
The determination of the conditions for the boundary line between
the superconducting and the insulating phase then reduces to the
explicit evaluation of $\chi_{ij}(\tau,\tau^{'})$. The result for
that has been first obtained in Ref. [48], and we sketch  below
the main steps of this derivation.

In determining  $\chi_{ij}(\tau,\tau^{'})$  it has turned
out to be essential to treat the phase variable $\phi_i$ in a compact form.
In order to separate the imaginary time evolution of the
phase into a periodic part $\bar \phi_i(\tau)$ and an non-periodic
part, one introduces the following parameterization:
\begin{eqnarray}
\phi_i(\tau)=\bar{\phi}_i(\tau)+ \frac{2\pi n_i \tau}{\beta}
\end{eqnarray}
with $\bar{\phi}_i(0)=\bar{\phi}_i(\beta)$ and $n_i$ being an integer which
counts how many times the phase winds over an angle which is a multiple of
$2\pi$.

The use of such a relation allows to express the sum over all
$\phi_i$ as an integration over $\bar{\phi}_i$ plus a sum over all the
possible integer values of the winding number $n_i$. After
performing a number of suitable algebraic operations\cite{Brud92}, one ends
up with the following expression for the two-time phase correlator:
\begin{eqnarray}
&&\chi_{ij}(\tau) = \delta_{ij} \frac{\exp[-2 C_{ii}^{-1} |\tau|]}
{\sum_{\{n_i\}} \exp\left[ -\sum_{ij} 2\beta C_{ij}^{-1} N_i N_j \right]}\times
\nonumber \\
&& \sum_{\{n_i\}}  \exp \left[ -\sum_{ij} 2\beta C_{ij}^{-1} N_i
N_j - \sum_{k} 4 C_{ik}^{-1} N_k \tau \right],
\end{eqnarray}
\noindent where $N_i= q_i/2 +n_i$. This two-time phase
correlator is local in space and its time dependence follows an
exponential behavior, if one assumes that the static {\it offset
charge} $q_i$ has a distribution which is homogeneous in
space. In the general case, of a dynamic {\it offset charge} $q_{i}(\tau)$,
this time dependence will be modified and can lead to qualitative changes
in the nature of the SIT.

Having obtained the expression for the local two-time  phase correlator in
the time and space representation, we  can now express the effective
Ginzburg-Landau free energy functional in the Fourier space in the
following way:
\begin{eqnarray}
F_{\psi}=\frac{1}{\beta L} \sum_{n,k} \psi_k^*(\omega_n)\left[
\alpha_k^{-1} -\chi_{k}(\omega_n)] \psi_{k}(\omega_n) \right]\, .
\end{eqnarray}
Expanding  the inverse  matrix $\alpha_{ij}^{-1}$ in the form
$\alpha_k^{-1}=(1/z)+k^2 (a^2/z^2)+...$, this free-energy functional
finally is written as:
\begin{eqnarray}
F_{\psi}=\frac{1}{\beta L} \sum_{k,n} \left[\frac{2}{z E_J}
-\chi_0 + a k^2+b \omega_n^2+... \right]
|\psi_k(\omega_n)|^2.\nonumber \\
\end{eqnarray}

The transition line is then given by
the condition that the coefficient of the quadratic term vanishes
in the limit of vanishing $k$ and $\omega$, that is:
\begin{eqnarray}
1-\frac{z E_J}{2} \chi_0(0)=0 \,. \label{boundary}
\end{eqnarray}

For a quantitative analysis, one has to determine the explicit zero
frequency limit of the two-time phase correlator. As mentioned
above, the inclusion of the time dependent {\it offset-charge} coming from
the fluctuations of the bosonic density can modify the low
frequency behavior of the local phase correlations.

In order to get a first insight into the underlying physics at
play here we evaluate the two time phase correlator in the limit
of a purely local {\it capacitance} (the so called {\it self-charging}
limit), by keeping only the on-site part in the original structure
of the {\it capacitance} matrix. Under those conditions the
evaluation of the two time phase correlator at finite frequency
gives:
\begin{eqnarray}
\chi_{ii}(\omega_n)=\frac{1}{Z_0} \sum_{\{n_i\}} F[n_i] \left(
\frac{1}{C_{ii}^{-1}}-\left[ i \omega_n -4
C_{ii}^{-1} n_i \right] \right)\nonumber \\
\end{eqnarray}
with $F[n_i]=\exp[-\sum_i 2 \beta C_{ii}^{-1} n_i^2]$ and
$Z_0=\sum_{\{n_i\}} F[n_i]$.

With these expressions we finally can cast Eq. \ref{boundary} in the form
\begin{eqnarray}
1- \frac{z E_J}{4 E_C} \frac{ \sum_{n} \frac{\exp[-4 \beta E_C
n^2]}{1-4 n^2}}{\sum_{m} \exp[-4 E_C \beta m^2]}=0 \,,
\end{eqnarray}
expressed in terms of the {\it Josephson coupling}  $E_J$ and the
{\it charging energy} $E_C=C_{ii}^{-1}/2$, as given by the
Eqs.(\ref{selfcapac},\ref{joseph},\ref{mutualcap}).
\begin{figure}[b]
\centerline{\psfig{figure=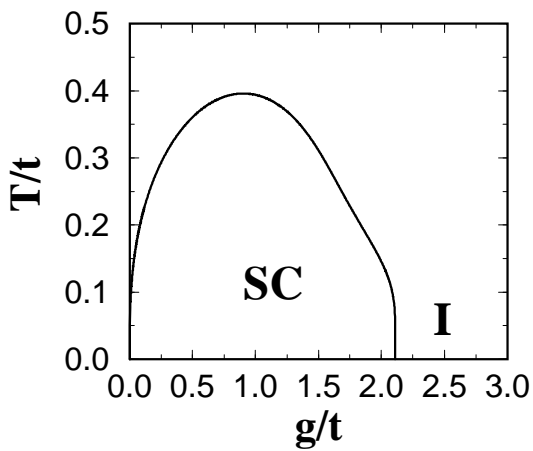 ,width=7.cm}} \caption{Phase
diagram as a function of $g/t$ describing the boundary between an
insulating(I) and superconducting(SC) phase with long range phase
coherence for the case of fully symmetric limit ($\Delta_B=0,
n_{F\uparrow} = n_{F\downarrow} = 2n_B =1$).}
\label{fig4-phasediagPI1}
\end{figure}

Given this defining equation for the boundary between the
superconducting and the insulating phase, we want to see now how
the intrinsic dependence of the Josephson and charging energy on
the exchange coupling and the density of bosons, manifests itself
in the competition between phase and boson density degrees of
freedom. In the loop expansion given in the previous section for
the BFM, we have obtained the amplitude of the intersite phase
coupling and the {\it charging} effect as a function of
temperature, the ratio $g/t$ and the bosonic density. It is the
non trivial dependence of those coupling constants on the
parameters of the original microscopic Hamiltonian of the BFM
which leads to the  interesting features in the competition
between the phase and boson density fluctuations. Thus, upon
increasing the pair exchange coupling it is expected that the {\it
Josephson} coupling is reduced while the {\it charging} energy is
increased due to the local transfer between fermions pairs and
bosons as well as due to the single particle hopping of fermions
between nearest neighbor sites.

We determine the critical line for two different cases in order to
highlight the role played by, on the one hand, the coupling and, on the
other hand, by the effective
boson density. In Fig. \ref{fig4-phasediagPI1}, we illustrate the
transition line $T_{\phi}$, separating a phase coherent state from
a phase disordered one, as a function of the coupling strength and
the effective boson density $n_B$. The evolution of the transition
line is non monotonic as a function of $g/t$  and goes through a
maximum at $g_{max}\sim t$. The quantum critical point, where the
SIT occurs, is given by $g_{crit} \sim 2 t$. The critical behavior
close to the transition is that of an XY model in d+1 dimensions.

\begin{figure}[t]
\centerline{\psfig{figure=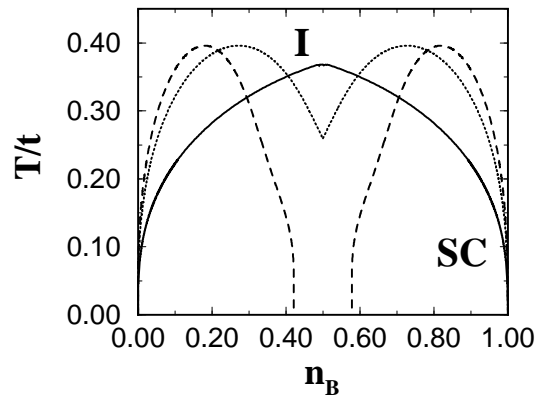 ,width=7.cm}} \caption{Phase
boundary lines between the superconducting(SC) and insulating(I)
state at different values of the coupling and upon  varying the
average density of bosons. Solid, dotted and dashed line stand for
values of the coupling constants for which
$g<g_{max}$,\,$g_{crit}>g>g_{max}$ and $g>g_{crit}$,
respectively.} \label{fig5-phasediagPI1}
\end{figure}

More interesting still is the behavior of $T_{\phi}$ as a function of
the $n_{B}$. The variation of $T_{\phi}$ is qualitatively
different for the different parameter regimes: a) the  weak
coupling case for $g<g_{max}$, b) the intermediate one with
$g_{max}<g<g_{crit}$ and  c) the strong coupling limit for
$g>g_{crit}$ (see Fig. \ref{fig5-phasediagPI1}). Going from the
limit $a$ to $c$ we find that with increasing $n_B$ the critical
line decreases to zero, goes through a maximum starting at a
finite $T^{\phi}(n_B=1)$ and finally shows a SIT at a critical
density.

%

\section{Role of Berry phase term: an intrinsic Magnus force on vortices}

In the preceding section we have analyzed the basic physics
resulting from the effective action by neglecting, (i) the
influence of any feedback between the density and amplitude
fluctuations (included in the field $l[\theta_i]$) on the phase
dynamics, and (ii) the contribution from the Berry phase term,
responsible for the correct quantization of the pseudospin
variable, which is given by the integral over all the possible
paths of $i s (1-cos[\theta_i])
\partial_{\tau} \phi_i$. In this section, we discuss
the consequences of the presence of such a Berry phase term in the
case where the phase action has a vortex solution. The existence
of a vortex solution is assured for 2D systems where the phase
correlations are described by a XY type dynamics. We shall show
here that the Berry phase term will produce an intrinsic Magnus
Force on the vortex which is analogous to the Lorentz force for a
charged particle, whose effective magnetic field  depends on the
spatial distribution of the $\theta$ field. Using the
correspondence with the bosonic pseudospin variable, this implies
a relation with the spatial density distribution of the bosons.
The Magnus force has been widely discussed in the context of
normal BCS type superconductors where it has been shown that it
arises from the Berry phase caused by the adiabatic motion of a
vortex along a closed loop coming back to its starting
position\cite{Ao93}. The adiabatic vortex motion on a loop in the
superconducting state turns out to be affected by an effective
magnetic field generated by the supercurrent arising from the
gradient of the phase which encircles such a vortex. In the BFM
scenario discussed here, we find that on top of the usual
contribution due to the superfluid electrons, there is an
intrinsic Berry phase term which will generate such a Magnus force
and which has an intensity proportional to $2n_B-1$, where $n_B$
is the bosonic density. This Magnus force, arising solely from the
quantum nature of the pseudospin variable, is clearly independent
on any superconducting state, and does not require a coherent
superfluid current induced by the presence of the vortex itself.
Moreover, since $n_B$ can be controlled externally, one has the
possibility to tune the strength of the effective magnetic field
acting on the vortex, and hence to alter its dynamical properties.
This will result in a possible measurable effect on transport
coefficients, such as the Hall coefficient, resistance, etc.

To be more explicit and following the procedures used in different
approaches treating with the Magnus force problem, let us assume
that one has a vortex centered at the position $R$. Let us
furthermore consider that we are in the continuum and at zero
temperature so that the Berry phase term in the Lagrangian now
reads $L_{B}=\int d^2 r cos[\theta(r,t)]
\partial_{t} \phi(r,t)$. The contributions linear in
$\partial_{t} \phi(r,t)$ and having time independent coefficients
do not contribute to the dynamics. Moreover, since we consider
that the phase $\phi$ and the variable $\theta$ are linked to a
vortex solution centering at $R$, it is judicious to introduce a
new relative variable $r-R$, which defines the intrinsic position
dependence of the variables with respect to the vortex center.

Starting from the above given expression for the Lagrangian
$L_{B}$ and making a simple change in the time derivative we can
rewrite it in the following form:
\begin{eqnarray}
L_{B}&&=\int d^2 r cos[\theta(r-R,t)]
\partial_{t} \phi(r-R,t)\\
&&=\frac{d {\bf R}}{dt}\int d^2 r cos[\theta(r-R,t)] {\bf \nabla}_r
\phi(r-R,t).
\end{eqnarray}

At this point, one recognizes that the term $\int d^2 r
cos[\theta(r-R,t)] {\bf \nabla}_r \phi(r-R,t)$ plays the role of
an effective vector potential ${\bf A}_{B}$ and $\frac{d {\bf
R}}{dt}$ represents the velocity of the vortex. This
Lagrangian is thus equivalent to that of a charged particle in
presence of an effective magnetic field in the z-direction given
by ${\bf B}={\bf \nabla} \times {\bf A}$. ${\bf B}$ hence creates
a Magnus force ${\bf F}_M = -(\frac{d {\bf R}}{dt} \times
\widehat{{\bf z}}) 2 \pi cos[\theta_0]$ which is transverse and
proportional to the vortex velocity and whose strength is linked
to the magnitude of this effective  magnetic field $B_z =-2 \pi
cos[\theta_0]$. Its magnitude is given by  the asymptotic value of
the background bosonic density at large distance $cos[\theta_0]$
for $r\rightarrow \infty$.

In case of a 2D superconductor described by the BFM scenario, the usual
effective magnetic field  $-2 \pi\rho_s$ arising from the supercurrent of
the electrons circling the vortex has thus to be supplemented by the above
mentioned contribution $B$ (intrinsic to such a scenario) and permits in
principle to modulate the total Magnus force upon changing the density of the
bosons. Clearly, a full description of such an eventuality, which is beyond
the scope  of the present analysis, has to be studied in more detail and has
to include the derivation of the effective action in presence of magnetic
field such as to determine the superfluid density of the electrons $\rho_s$.

At this stage we simply want to point out that the sign of the
magnetic field arising from the Berry phase term in the action of
the BFM changes if one tunes the bosonic density between zero and
unity. In terms of the variable $\cos[\theta]$ this implies a
variation in the range $[-1,1]$. In other words, the sign of the
intrinsic Berry phase induced Magnus force will change at
$n_B=1/2$ when going from the limit of small density of bosons to
high density. This extra contribution to the Magnus force has to
be added of course to the conventional one known for standard
superconductors.

%
\section{Comparison with negative-U and Bose-Hubbard models}

A frequently asked question concerns the qualitative differences
which exist between the BFM and similar scenarios such as the
negative-U Hubbard model and pure bosonic systems such as the
Bose-Hubbard model and Josephson junction arrays.

Let us start with a comparison of the BFM and the negative-U
Hubbard model which has been studied in a great variety of
different approaches and discussed especially in connection with
the BCS-BEC crossover. The main issue of such a comparison is the
occurrence or not of a quantum critical point in the negative-U
Hubbard scenario at a finite value of the coupling $U$ which
introduces the pairing among the electrons. As we have shown in
this paper, in the BFM there exists a critical value for the
exchange coupling $g/t$ responsible of the pairing in the
fermionic subsystem above which the system undergoes a SIT. We
have seen that within a path integral formalism, one can extract
an effective action describing such a transition as being
primarily due to the interplay between the \underline{local}
quantum pair exchange (between a boson and a fermion pair) and the
\underline{non-local} inter-site pair hopping leading to a phase
dynamics originating from the itinerancy of the fermions. Due to
the non-trivial dependence of the strength of these two competing
mechanisms on the microscopic parameters of the BFM, there occurs
a SIT for a finite density as well as finite exchange coupling
$g/t$. For the negative-U Hubbard scenario this is not the case
and the SIT does not occur at any finite value of the ratio
$U/t$\cite{Schneider}. This model merely describes a continuous
cross-over between a  BCS superconductor and a BEC of tightly
bound electron pairs as U is increased from 0 to $\infty$.

In order to better understand this difference between the BFM and
the negative-U Hubbard model, let us consider two scenarios in
equivalent situations, i.e., the half filled band case for the
negative-U Hubbard model and the fully symmetric case for the BFM
(with the bosonic level lying in the middle of the fermionic band
such that both, the fermionic band as well as the bosonic level
are half occupied). We then address the question how, in the
strong coupling limit, the pair hopping is generated out of the
basic configurations of states and processes which contribute in
this regime.

For the case of the negative-U Hubbard model with fermions
interacting via a local attractive potential, the ground state
wave function in the limit of $U\rightarrow \infty$ and at
half-filling is highly degenerate and is composed of all the
possible configurations comprising equal distribution of zero and
doubly occupied site. In this limit, one can perform a mapping of
this model on the hard-core boson model described by the
Hamiltonian:
\begin{eqnarray}
H_U=-\sum_{\langle ij \rangle} \frac{2 t^2_{ij}}{U} b^{+}_i
b_{j}+\sum_{\langle ij \rangle} \frac{2 t^2_{ij}}{U} n_i
n_{j}-\tilde{\mu}\sum_i n_i.
\end{eqnarray}
$b_i^{+}(b_i)$ and $n_i=b_i^{+} b_i$ stand for the creation
(destruction) operators of hard-core bosonic particles (tightly
bound electron pairs) and for their density operator,
respectively. Due to the presence of a coupling, which is
isotropic both in the boson hopping and in the charge interacting
channel, one has a superconducting state for any finite value of
the ratio $U/t$, and possibly a supersolid phase  due to the
symmetry of the charging interactions characterized by a
coexistence of diagonal and off-diagonal long range order. This
implies that the quantum critical point is strictly pushed to
$U\rightarrow \infty$.

In the BFM the phase space in the large $g/t$ limit is completely
different. In the fully symmetric case of this model (corresponding  to a total
density equal  twice the number of sites ($n_{tot}=2$)), the ground state
configuration is given by a wave function where all the bosons
are strongly coupled with pairs of fermions, thus resulting in a product state
of local bonding states:
\begin{eqnarray}
|\psi_0 \rangle=\prod_i \frac{1}{\sqrt{2}}
(\rho^{+}_{i}+\tau^{+}_i)|0 \rangle . \label{locstate}
\end{eqnarray}
This wave function can be viewed as a ferromagnetic Ising-type
state in the sense that it is made up of a ferro-type bonding
order and is not degenerate with respect to all the other possible
local configurations. Moreover, by construction it contains
bonding-bonding correlations on a long range (bonding solid), but
has zero phase correlation length. The latter can be shown by
evaluating the static correlation function for the bosons or,
equivalently,  the fermion pairs which gives $\langle \psi_0 |
\rho_i^{+} \rho_j | \psi_0 \rangle =0$ and $\langle \psi_0 |
\tau_i^{+} \tau_j | \psi_0 \rangle =0$ for any distance $|i-j|$.

Next, let us consider the low energy configurations which are
mixed into such a $g/t = \infty$ ground state when the kinetic energy
operator $H_t=\sum _{i\neq j,\:\sigma } t_{ij} (c_{i\sigma
}^{+}c_{j\sigma } +H.c.)$ is switched on. Applying $H_t$ to $|
\psi_0 \rangle$, there will occur states which are separated by an
energy gap with respect to $|\psi_0 \rangle$. The low energy
states to be considered as the relevant quantum mixed
configurations are of non-bonding nature, such as $C^+_{i
\sigma}|0\rangle = c^+_{i \sigma}|0\rangle$ and
$S^+_{i\sigma}|0\rangle = \rho^+_i c^+_{i\sigma} |0\rangle$. In
order to construct those states, let us to begin with consider a
local excitation on two adjacent sites $i,j$ given by $S^+_{i
\sigma}C^+_{j \, -\sigma} |0 \rangle$ in the background of bonding
states.  This configuration can be seen as a ferro-type order
interrupted by two domain walls of non-bonding/bonding nature.
Now let us consider the dynamics of such objects and
under which conditions they can be rendered itinerant in a way which
leaves the number of bonding configurations unchanged. The problem
is thus analogous to that of domain wall dynamics in an Ising type
systems. It so happens (a detailed discussion of this is beyond
the scope of the present study and will be given at a later stage
elsewhere), that the local degeneracy of the non-bonding states
will be removed by the action of the kinetic term. In this way it
induces a global lowering of their excitation energy due to the
dispersion of each domain wall which is of the form
\begin{eqnarray}
E_{DW}(k)=\frac{1}{2}[\epsilon_k+6 g\pm \sqrt{(\epsilon_k-2g)^2+
4t^2}]\,
\end{eqnarray}
\noindent and where $\varepsilon(k) =-2zt cos k$. Only after the
condensation of the domain wall like excitations in the presence
of the  bonding state background one can meet the  conditions for
setting up long range boson and fermion-pair phase correlations
.

This simple sketch of the nature of the excitations in the BFM
implies that the onset of phase coherence cannot be activated for
an arbitrarily small hopping amplitude since one has to overcome
the energy gap between the ground state and the manifold of
non-bonding states, which is of the order of $\sim 2g$. As we can
see from the expression for the dispersion of the domain walls,
this is achieved when $E_{DW}(k)$ becomes zero for $k=0$,  i.e.,
when $g/t=\frac{1}{4}(2+\sqrt{2+4 z^2})$. For z=2 this gives
$g/t\sim 2.0$, which determines when those defects-like
excitations become gapless and thus induce a proliferation within
the background of the bonding states. It is worth pointing out
that this value for the critical exchange coupling matches rather
well the value obtained from our functional integral approach
discussed in this paper (see Fig 3).

The picture which hence emerges for the BFM is very similar to
that of a XY model in a transverse field. There, the hopping term
is responsible of the XY dynamics, while the local boson-fermion
pair exchange provides the role of the effective transverse field.
This itself is already a strong indication that this model is of
different nature to that of the negative-U Hubbard scenario, for
which we know that it is, on the contrary, akin to an isotropic
Heisenberg model.

A further essential difference between the BFM and the negative-U
Hubbard model can be seen from their respective path-integral
formulations. When the path integral representation is considered
for this model\cite{Kopec}, the first step is to make use of a
Hubbard-Stratonovich transformation to rewrite the quartic
interaction term in a bilinear form, where the fermionic operators
are now coupled to random auxiliary fields. Limiting oneself to
purely superconducting order, the usual procedure for manipulating
such a bilinear action is to separate the complex auxiliary field
into its modulus and a pure phase part before performing an
expansion around the saddle point solution for the amplitude of
this auxiliary field. This way of proceeding allows then to
extract an effective action for the slow phase dynamics. This is
distinctly different from the functional integral representation
for the BFM which we have presented above and where from the very
beginning and throughout such a procedure the fermionic pair
fields are coupled to physically real bosonic modes which have
their own proper dynamics. As we have seen in section III, the
bosonic part of the action is treated within a coherent
pseudo-spin representation where the dynamics of the pseudospin is
parameterized by the time dependence of the spherical variables.
The role of the dynamics within the spherical representation of
the bosonic field and of the Berry phase term (which is a
consequence of the quantum interference in the local pseudospin
space) is a distinct feature of this BFM and presents specific
differences with respect to the negative-U model. Such differences
imply in particular that in the BFM case one has feedback effects
between the amplitude and the phase fluctuations which are totally
absent in an equivalent description of the negative-U Hubbard
model. Such features are important because of intrinsic
dissipation effects that eventually can change the nature of the
transition and possibly are relevant for the emergence of a
"bosonic metal" in proximity of the SIT, a feature which seems to
be outside the framework of the negative-U Hubbard model.

Let us conclude this comparison of the BFM with similar models
with a brief discussion on the Bose-Hubbard model and Josephson
junction arrays. The main aspect which emerges as a common
denominator in all those models, at least as far as the phenomenon
of the SIT is concerned, is that the mechanism which is
responsible for the degrading of the phase coherence is analogous
and originates from the competition between the phase and charge
degrees of freedom. In the Bose-Hubbard model this manifests
itself as a competition between the boson hopping and their charge
repulsion, while in the Josephson junction array scenario it
appears as an interplay between the Josephson tunnelling amplitude
and the charging energy.

In spite of this, at first sight, similarity between the BFM and
those scenarios, we would like to stress that in a system where
the dynamics is described by the BFM scenario, the interplay
between the phase and amplitude fluctuations is intrinsically
related to the coupling between the fermionic and bosonic degrees
of freedom. This introduces amplitudes of the different processes
at work in form of a non-trivial dependence on the microscopic
parameters of the starting Hamiltonian. Furthermore, as we have
seen in the above discussion of the effective action of the BFM,
this model contains features which go beyond those which
characterize the pure Josephson-type dynamics and which arise from
the peculiar feedback between the fluctuations of the bosonic
density (or amplitude pair field) and fluctuations of the phase.
Last not least, the appearance of an intrinsic Berry phase-term in
the BFM scenario together with the effect of dissipation due to
the fermionic dynamics, can give rise to an unconventional
phenomenology when topological phases play a role, and especially
in the presence of vortices.
\section{Conclusion}

In this paper we examined the nature of a superconductor -
insulator transition in a system of localized bosons and itinerant
fermions coupled together via a pair exchange term. An effective
action was derived from such a microscopic model, which, after
integrating out the fermionic fields, could be phrased in terms of
amplitude and phase fluctuations of the bosons. In order to make
the presentation more familiar we discussed the action in a
terminology frequently employed in connection with the study of
Josephson junction arrays. We stress that our system does not
necessarily imply any charged fermions and bosons.

Considering the phase-only part of the effective action it is
fully equivalent to the quantum phase model for the Josephson
junction arrays, discussed in terms of: (i) a {\it Josephson
coupling} term, (ii) a {\it charging or capacitance} term and
(iii) an {\it offset charge} term. Equivalent to that in our
scenario is: (i) a boson hopping term, (ii) a term which takes
into account the reduction in hopping amplitude due to a
fluctuating local boson density arising from the intrinsic on-site
exchange coupling between the bosons and the fermions and (iii) a
chemical potential term controlling the bosonic concentration. New
in the present study within already the lowest (phase-only)
approach to the boson-fermion system is the intricacy of the
dependence of the effective Josephson coupling, the capacitance
term and the off-set charge term on the parameters of the initial
Hamiltonian, i.e., the exchange coupling and the density of bosons
(or else the total particle density). It turns out that within the
lowest (phase-only) contribute a superconductor to insulator
transition can not only be triggered by a change in the exchange
coupling but also by a variation of the boson density. The latter
presents evidently a certain interest from the experimental point
of view and can possibly be tested in such transition occurring in
optical lattices for ultracold fermi gases with Feshbach resonance
pairing.

Apart from the phase-only part of the effective action we
established the existence of an intrinsic Berry phase term, which
arises from the hard core nature of the bosons and gives rise to
an additional Magnus force when the system is such that
topological ground state configurations like vortices are
stabilized. This again is of potential interest for experiments
since in principle one can change the sign of the Magnus force
upon changing the density of bosons and thus deviating the motion
of a vortex from one direction into the opposite one. These very
preliminary results will be dealt with in greater detail in some
future studies.

A further property of the Berry phase term is that it gives rise
to a bilinear coupling term between the phase and amplitude
fluctuations, and is hence much more direct and relevant than
similar terms in, for instance, scenarios based on the negative-U
Hubbard model where they occur only at a much higher order in a
corresponding loop expansion of the trace in the effective action.
This again merits to be investigated in some detail with the aim
to study the dissipation introduced by such amplitude phase
coupling and its effect on the nature of the transition, in view
of exploring the possibility of an intermediary bosonic metallic
ground state.

Finally, we have dealt with a frequently asked question concerning
the differences between the negative-U Hubbard scenario - mainly
studied in connection with the BCS-BEC cross-over and the
presently studied boson-fermion model. Far from being able to give
a complete account for the major differences we found mainly two
aspects which distinguish the physics of these models on a
qualitative and robust level. The one is that in the negative-U
Hubbard model a superconductor - insulator transition can not take
place at any finite coupling U, nor can such a transition be
triggered by the change in particle concentration. The second
point is that the ground states in the strong coupling limit of
the two models are quite different: a highly degenerate ground
state for the negative-U Hubbard model with excitations being
controlled by an isotropic Heisenberg model when the hopping term
is switched on. Contrary to that, for the boson fermion model the
ground state is non-degenerate (corresponding to a ferro
pseudo-magnetic Ising type system of singlets formed by bosons and
fermion pairs), which, after switching on the fermion hopping
term, gives rise to propagating domain like structures. The
topological structures appearing in the ground state might have
measurable consequences in the transport properties near the
superconductor-insulator transition.

This and the other preliminary studies mentioned above will
require a detailed analysis and will be discussed in future.

\end{document}